# Modified-Bloch Equation Based on Fractal Derivative for Analyzing PFG Anomalous Diffusion

**Guoxing Lin***

*Carlson School of Chemistry and Biochemistry, Clark University, Worcester, MA 01610*

*September 2017*

ABSTRACT: A modified-Bloch equation based on the fractal derivative is proposed to analyze pulsed field gradient (PFG) anomalous diffusion. Anomalous diffusion exists in many systems such as in polymer or biological systems. PFG anomalous diffusion could be analyzed based on the fractal derivative or the fractional derivative. Compared to the fractional derivative, the fractal derivative is simpler, and it is faster in numerical evaluations. In this paper, the fractal derivative is employed to build the modified-Bloch equation that is a fundamental method to describe the spin magnetization evolution affected by fractional diffusion, Larmor precession, and relaxation. An equivalent form of the fractal derivative is proposed to convert the fractional diffusion equation, which can then be combined with the precession and relaxation equations to get the modified-Bloch equation. This modified-Bloch equation yields a general PFG signal attenuation expression that includes the finite gradient pulse width (FGPW) effect, namely, the signal attenuation during field gradient pulse. The FGPW effect needs to be considered in most clinical MRI applications, and including FGPW effect allows the detecting of slower diffusion that is often encountered in polymer systems. Additionally, the spin-spin relaxation effect can be analyzed, which provides a broad view of the dynamic process in materials. The modified-Bloch equation based on the fractal derivative could provide a fundamental theoretical model for PFG anomalous diffusion.

**Introduction**

Over the last few decades, significant amounts of anomalous diffusion studies have been carried out in a variety of systems such as in polymer or biological systems.[1,2,3,4,5,6] Anomalous diffusion can be detected by pulsed field gradient (PFG) diffusion experiments.[7,8,9,10,11,12,13] PFG technique is a noninvasive tool to monitor diffusion.[14,15,16,17,18,19] It has broad applications in Nuclear Magnetic Resonance (NMR) to study the transport property of the real material, and in Magnetic Resonance Imaging (MRI) to build medical images. In PFG experiments, the detected signal intensity depends upon the diffusion behavior and the applied gradient pulses; the faster diffusion induced a larger signal attenuation. The analysis of PFG anomalous diffusion is much complicated than normal diffusion because anomalous diffusion has a non-Gaussian probability distribution and a nonlinearly time-dependent mean square displacement. There are sophisticated PFG theories for normal diffusion,[17,18] while, PFG theories for anomalous diffusion are developed relatively recently,[7,9,12,13] and it is insufficient to analyze PFG anomalous diffusion. The PFG anomalous diffusion theory could explain the non-monoexponential attenuation of large b-value data in non-Gaussian diffusion that is hard to be analyzed by PFG normal diffusion theory.[20] Additionally, for the curvilinear diffusion existing in macromolecule systems,[9,21,22] the analysis by PFG anomalous diffusion theory could give more

*To whom correspondence should be addressed. Email: glin@clarku.edu.

accurate diffusion domain size.[23,24] Furthermore, the analysis by PFG anomalous diffusion theory can give the time and space derivative parameters.[25] These two parameters are related to the diffusion jump length and time distributions[2] that are determined by material dynamics properties. To better analyze non-Gaussian diffusion, many efforts have been devoted to developing PFG anomalous diffusion theoretical methods.[7,9,10,11,12,13, 26, 27,28,29]

Although PFG anomalous diffusion can be described by the fractional derivative,[12,13,26] the recently developed fractal derivative has been adopted to describe PFG anomalous diffusion as well.[13, 23, 26, 27,29,30] The fractal derivative is developed based on the concepts of the Hausdorff derivative.[31,32] It is different from fractional derivative. The fractional derivative is a global derivative operator, while the fractal derivative is a local operator.[32] The fractal derivative gives a stretched exponential function based PFG signal attenuation expression,[13,23,29,30] which is consistent with other reported results where approximate or empirical stretched exponential functions has been proposed for PFG anomalous diffusion in polymer and biological systems.[7,8,10,11, 33] The fractal derivative is faster in numerical evaluations than the fractional derivative,[29] which is an advantage in MRI diffusion imaging where the data size can be large. Additionally, reference 29 reported that the fractal derivative could be a more natural way to link sub-voxel tissue composition with the PFG signal attenuation in MRI. Therefore, the fractal derivative has its advantages in analyzing the PFG anomalous diffusion, and it is important to develop a fundamental PFG theoretical treatment based on the fractal derivative.

Here, the fractal derivative is employed to build a modified-Bloch equation, which is one of the most fundamental PFG theoretical methods. The modified-Bloch equation[34,35] describes the spin magnetization evolution under the co-effect of anomalous diffusion, Larmor precession, and relaxation processes. Although various groups have proposed the modified-Bloch equations based on the fractional derivative,[12,36,37] the fractal derivative has not been used to build the modified-Bloch equation. Based on the fractal derivative,[31,32] the fractional diffusion equation has a time fractal derivative $\frac{\partial}{\partial t^\alpha}$, whose time dimension is different from $\frac{\partial}{\partial t}$ in the precession and the relaxation equations. The different time dimensions prevent the direct combination of the fractional diffusion equation with the precession and relaxation equations. This hurdle can be circumvented by rewriting the time fractal time derivative $\frac{\partial}{\partial t^\alpha}$ as $\frac{1}{\alpha t^{\alpha-1}}\frac{\partial}{\partial t}$; the factor $\frac{1}{\alpha t^{\alpha-1}}$ can be moved to the other side of the fractional diffusion equation, which can then be used to build the modified-Bloch equation. The obtained modified-Bloch equation can analyze PFG signal attenuation including the finite gradient pulse width (FGPW) effect, namely the attenuation during the gradient pulse period.[17,18] The signal attenuation during a short gradient pulse could be neglected in PFG experiments, which is referred to as the short gradient pulse (SGP) approximation,[17,18] nevertheless, the routine gradient pulse in clinical MRI is sufficiently long, and a long gradient pulse can measure slower diffusion under the same maximum gradient strength. It is thus important to take into account of the FGPW effect. The spin-spin relaxation effect[8] can be derived from the modified-Bloch equation as well. The modified-Bloch equation based on the fractal derivative provides a fundamental way to analyze PFG anomalous diffusion.

**Theory**

Modified-Bloch equation has been a primary PFG theoretical treatment for normal diffusion,[17,18] which can be viewed as a specific case of anomalous diffusion. The modified-Bloch equation for normal diffusion is[35]

$$\frac{\partial}{\partial t} M_{xy}(z,t) = D \frac{\partial^2}{\partial z^2} M_{xy}(z,t) - i\gamma g(t) z M_{xy}(z,t) - \frac{M_{xy}(z,t)}{T_2}, \tag{1}$$

where $M_{xy}(z,t) = M_x(z,t) + iM_y(z,t)$ is the magnetization, $z$ is the position, $g(t)$ is the time-dependent gradient, and $T_2$ is the spin-spin relaxation time constant. For simplicity, only one-dimensional diffusion along $z$ direction is considered here. Equation 1 is a linear combination of simultaneous processes: diffusion, Larmor precession, and relaxation. In a rotating frame rotating around a magnetic field at angular frequency $\omega = -\gamma B_0$ where $B_0$ is the exterior magnetic field, and $\gamma$ is the gyromagnetic ratio, the spin Larmor precession can be described by[17,18]

$$\frac{\partial}{\partial t} M_{xy}(z,t) = -i\gamma g(t) \cdot z M_{xy}(z,t). \tag{2}$$

The relaxation of the transverse component of magnetization can be described as[38]

$$\frac{\partial}{\partial t} M_{xy}(z,t) = -\frac{M_{xy}(z,t)}{T_2}. \tag{3}$$

The one-dimensional normal diffusion can be described by[39]

$$\frac{\partial}{\partial t} M_{xy}(z,t) = D \frac{\partial^2}{\partial z^2} M_{xy}(z,t), \tag{4}$$

where $D$ is the diffusion coefficient. For normal diffusion, equations 2-4 can be linearly combined to build the modified-Bloch equation, because they have the same time derivative $\frac{\partial}{\partial t}$. However, for anomalous diffusion, the fractal time derivative is significantly different from $\frac{\partial}{\partial t}$. Based on the fractal derivative which is developed based on the concepts of the Hausdorff derivative, the time-space fractional diffusion equation is[31,32]

$$\frac{\partial}{\partial t^\alpha} M_{xy}(z,t) = D_f \frac{\partial}{\partial z^{\beta/2}} \left( \frac{\partial M_{xy}(z,t)}{\partial z^{\beta/2}} \right), \tag{5}$$

where $D_f$ is the fractional diffusion coefficient with units $m^\beta / s^\alpha$, and the fractal derivative is defined as[31,32]

$$\frac{\partial P^\beta}{\partial t^\alpha} = \lim_{t_1 \to t} \frac{P^\beta(t_1) - P^\beta(t)}{t_1^\alpha - t^\alpha}, 0 < \alpha, 0 < \beta. \tag{6}$$

The fractal time derivative $\frac{\partial}{\partial t^\alpha}$ is distinct from $\frac{\partial}{\partial t}$, which makes it difficult to combine the anomalous diffusion equation with precession and relaxation equations. However, by rewriting the time fractal time derivative $\frac{\partial}{\partial t^\alpha}$ as $\frac{\partial}{\alpha t^{\alpha-1} \partial t}$, the fractional diffusion equation (5), can be rewritten as

$$\frac{\partial M_{xy}(z,t)}{\partial t} = \alpha t^{\alpha-1} D_f \frac{\partial}{\partial z^{\beta/2}}\left(\frac{\partial M_{xy}(z,t)}{\partial z^{\beta/2}}\right). \tag{7}$$

Equations 2, 3 and 7 have the same time derivative $\frac{\partial}{\partial t}$, and they can be combined to give a modified Bloch equation for anomalous diffusion:

$$\frac{\partial M_{xy}(z,t)}{\partial t} = \alpha t^{\alpha-1} D_f \frac{\partial}{\partial z^{\beta/2}}\left(\frac{\partial M_{xy}(z,t)}{\partial z^{\beta/2}}\right) - i\gamma g(t)z M_{xy}(z,t) - \frac{M_{xy}(z,t)}{T_2}. \tag{8}$$

When $\alpha = 1, \beta = 2$, eq 8 reduces to eq 1, the modified-Bloch equation for the PFG normal diffusion. In a homogenous system, the magnetization induced by the gradient field can be written as[17,18]

$$M_{xy}(z,t) = S(t)\exp(-iK(t)z), \tag{9}$$

where

$$K(t) = \int_0^t \gamma g(t)dt' \tag{10}$$

is the wavenumber. By substituting $M_{xy}(z,t) = S(t)\exp(-iK(t)z)$ into the modified-Bloch equation, eq 8, and applying $\frac{\partial}{\partial z^{\beta}}\exp(-iK(t)z) = -K^{\beta}(t)\exp(-iK(t)z)$[31,32] where oftentimes $K(t) \geq 0$, we have

$$\frac{\partial S(t)}{\partial t} = -\left(\alpha t^{\alpha-1} D_f K^{\beta}(t) + \frac{t}{T_2}\right)S(t). \tag{11}$$

The solution of eq 11 is

$$S(t) = \exp(-\frac{t}{T_2})\exp\left(-\int_0^t D_f K^{\beta}(t')dt'^{\alpha}\right). \tag{12}$$

If the relaxation effect is neglected, eq 12 reduces to

$$S(t) = \exp\left(-\int_0^t D_f K^{\beta}(t')dt'^{\alpha}\right), \tag{13}$$

which reproduces the signal attenuation expression obtained by two other methods, the effective phase shift diffusion equation method[13] and the instantaneous signal attenuation method[26]. Under SGP approximation, for the pulsed gradient spin echo (PGSE) sequence or the pulsed gradient stimulated-echo (PGSTE) sequence as shown in Figure 1, eq 13 reduces to

$$S(t) = \exp\left[-(\gamma g \delta)^{\beta} D_f \Delta^{\alpha}\right]. \tag{14}$$

Additionally, in the first period $0 < t \leq \delta$ of PGSE or PGSTE experiments, when $\beta = 2$, $K^2(t) = (\gamma g)^2 t^2$ and signal attenuation equation, eq 12 can be reduced to $\frac{\partial S(t)}{\partial t} = -\left( D_f \alpha (\gamma g)^2 t^{\alpha+1} + \frac{t}{T_2} \right) S(t)$. That is consistent with the result $\frac{\partial S(t)}{\partial t} = -Dg^2 t^{2-\nu} S(t), \nu = 1-\alpha$, obtained by spectral function and echo damping method.[40]

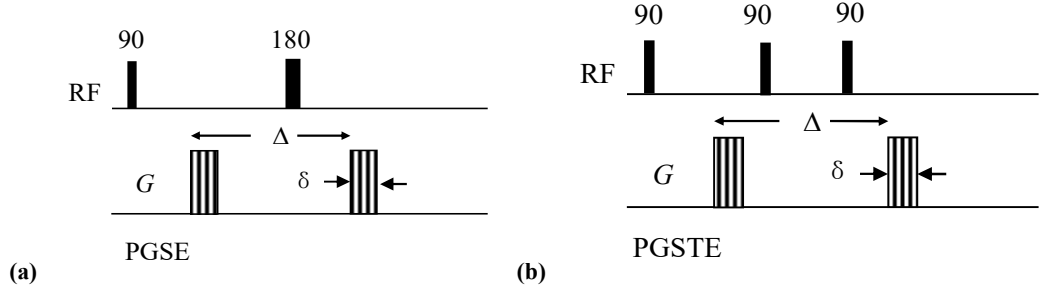

**Figure 1.** Two fundamental PFG pulse sequences: (a) pulsed gradient spin echo (PGSE) sequence, (b) pulsed gradient stimulated-echo (PGSTE) sequence. The gradient pulse width is $\delta$, and the diffusion delay is $\Delta$.

**Results and discussion**

This paper proposes the modified-Bloch equation based on the fractal derivative for PFG anomalous diffusion. From the modified-Bloch equation, the PFG signal attenuation expression is a stretched exponential function, $\exp(-\frac{t}{T_2}) \exp\left(-\int_0^t D_f K^\beta(t') dt'^\alpha\right)$. Neglecting the relaxation term $\exp(-\frac{t}{T_2})$, the PFG signal attenuation $\exp\left(-\int_0^t D_f K^\beta(t') dt'^\alpha\right)$ is the same as those obtained by the effective phase shift diffusion method[13] and the instantaneous signal attenuation method[26]. The effective phase diffusion equation method describes the phase evolution process, while the instantaneous signal attenuation method considers the signal average of diffusion. The results here also agree with two other approximation methods: the auto-correlation function method[7] and the modified-Gaussian phase shift distribution method[27]. Although the auto-correlation method[7] is a very successful pioneer method, it has considered only the time fractional diffusion with $\beta$ equaling 2, where its results are the same as that from the modified-Gaussian phase shift distribution method. The current approach can handle the general anomalous diffusion $\{0 < \alpha, \beta \leq 2\}$, which includes the tine-fractional diffusion $\{0 < \alpha \leq 2, \beta = 2\}$, space-fractional diffusion $\{\alpha = 1, 0 < \beta \leq 2\}$, and normal diffusion $\{\alpha = 1, \beta = 2\}$. Figure 2 shows the comparison between the three methods mentioned above. Figure 2(a) shows the time-fractional diffusion results with $\beta = 2$, while Figure 2(b) shows the general fractional diffusion with $\beta \neq 2$. In Figure 2, the results from the modified-Bloch equation have better agreement with the other two methods at small $\delta/\Delta$ than at large $\delta/\Delta$. Additionally, the agreement also depends on the value of $\beta$.[27]

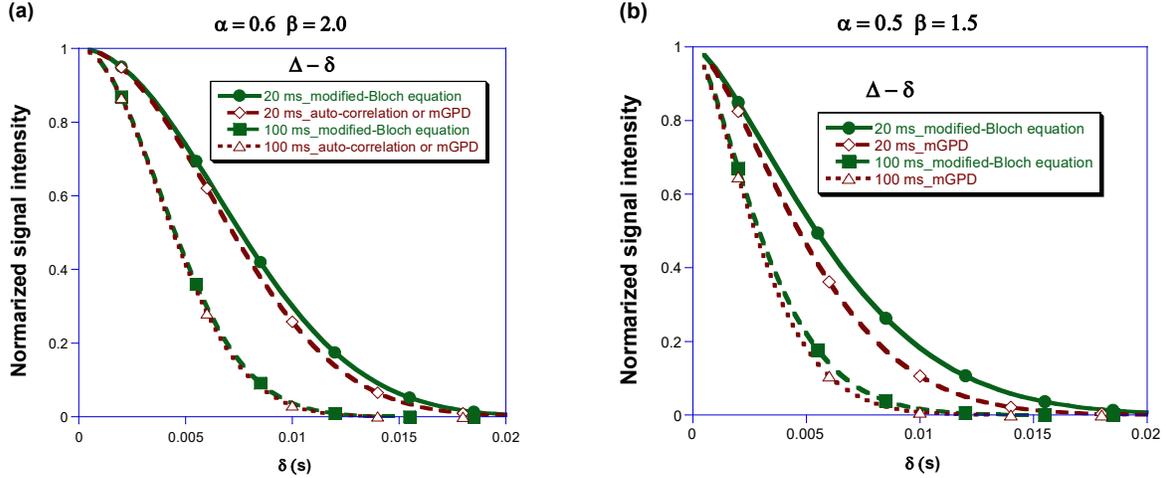

**Figure 2.** Comparison of the signal attenuation obtained by the modified-Bloch equation with that obtained by the modified-Gaussian approximation method in Refs. [15] and [17]: (a) $\alpha = 0.6, \beta = 2$, $D_f = 2.08 \times 10^{-10}$ m$^\beta$/s$^\alpha$, (b) $\alpha = 0.5, \beta = 1.5$, $D_f = 1.0 \times 10^{-7}$ m$^\beta$/s$^\alpha$. Other used parameters are $g$ equaling 0.1 T/m. $\gamma g \delta = 6 \times 2.6751 \times 10^4$ rad/m and $\gamma = 2.6751 \times 10^8$ rad/sT.

Determining the fractional derivative order parameters, $\alpha$ and $\beta$ is important for anomalous diffusion. $\alpha$ and $\beta$ are related with the diffusion jump time distribution and jump length distribution respectively, which could provide important dynamic information in material studies. Additionally, $\alpha$ and $\beta$ may play an important role in MRI imaging as they could potentially be used as contrast parameters for imaging.[25] These two parameters can be determined by PFG experiments. Palombo et al. have used PFG NMR to measure parameters $\alpha$ and $\beta$ by phenomenological signal attenuation formalisms for time-fractional diffusion and space-fractional diffusion.[33] Additionally, references 26-28 reported that $\alpha$ and $\beta$ can be determined by

$$\ln[\ln(S(0) - \ln S(t)] = \begin{cases} c_1 + \beta \ln(g), & \text{when } \delta \text{ and } \Delta \text{ are fixed} \\ c_2 + (\alpha + \beta) \ln(\delta), & \text{when } \Delta = \delta, \text{ and } g \text{ are fixed} \\ c_3 + \alpha \ln(\Delta), & \text{when } \delta \ll \Delta, \text{and } g \text{ are fixed} \end{cases} \quad (15)$$

where $c_1$, $c_2$ and $c_3$ are constants. Please refer to[26-28] for more detailed information.

The fractal derivative gives a stretched exponential function (SEF) based attenuation, while another fractional derivative operator, the fractional derivative gives a Mittag-Leffler function (MLF) based attenuation.[12,13,36] The comparison between stretched exponential attenuation and Mittag-Leffler function based attenuation is shown in Figure 3. At a small level of signal attenuation, the PFG signal attenuation based on the fractal derivative is close to the signal attenuation obtained by the fractional derivative, while at large attenuation, their difference becomes significant. Moreover, reference 13 reported that the stretched exponential attenuation attenuates faster than the Mittag-Leffler attenuation at subdiffusion, but it attenuates slower than the Mittag-Leffler attenuation at superdiffusion. It is also worth mentioning that at superdiffusion, the Mittag-Leffler function based signal attenuation from the fractional derivative model oscillates between positive and negative values at large attenuation regions,[13] which still cannot be clearly explained. However, the oscillating pattern does not appear in the stretched exponential attenuation obtained by the fractal derivative model.[13] Although the exact application scopes of the two models are still not clear,[31,32] these two derivative models have their own advantages, and they demonstrate complementary nature in analyzing PFG anomalous diffusion.

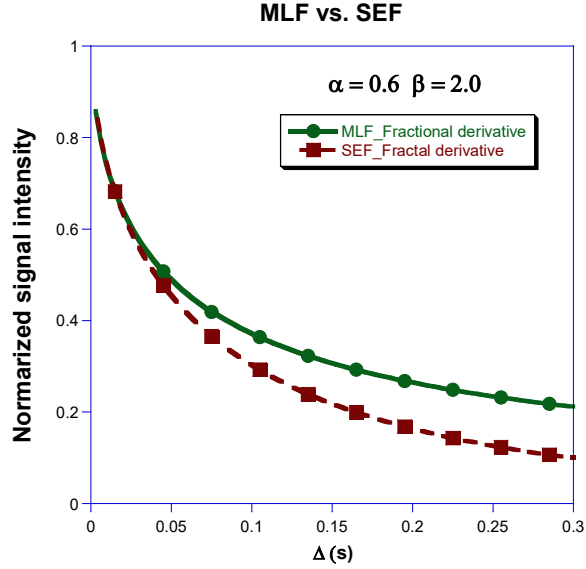

**Figure 3.** Comparison of the signal attenuation eq 14 obtained from the fractal derivative with $S(t) = E_\alpha\left[-(\gamma g \delta)^\beta D_f \Delta^\alpha\right]$ obtained from the fractional derivative.[37] The SGP approximation are used. The parameters used are $\alpha = 0.6, \beta = 2$, and $\gamma g \delta = 6 \times 2.6751 \times 10^4$ rad/m. Additionally, $D_f$ is $2.08 \times 10^{-10}$ m$^\beta$/s$^\alpha$ in the stretched exponential attenuation based on the fractal derivative, while it is $2.08 \times 10^{-10} / \Gamma(1+\alpha)$ m$^\beta$/s$^\alpha$ in the Mittag-Leffler function attenuation based on the fractional derivative.

The modified-Bloch equation can either be a differential equation or an integral equation. The modified-Bloch equation obtained based on the fractal derivative is a differential type equation, which is different from the integral type modified-Bloch equation obtained based on the fractional derivative.[12,36,37] However, the differential modified-Bloch equation, eq 8 can be transferred to an equivalent integral type of equation,

$$M_{xy}(z,t) = M_{xy}(z,0) + \int_0^t \left[\alpha t^{\alpha-1} D_f \frac{\partial}{\partial z^\beta} M_{xy}(z,t) - i\gamma g(t) z M_{xy}(z,t) - \frac{M_{xy}(z,t)}{T_2}\right] dt . \qquad (16)$$

Therefore, a modified-Bloch equation is built based on the combination of diffusion, precession, and relaxation processes at each instant of spin magnetization evolution. It can either be a differential equation, $\frac{\partial M_{xy}(z,t)}{\partial t} = \sum_i f_i(M_{xy}(z,t))$ , where $i$ stands for one of these processes, or an integral equation, $M_{xy}(z,t) = M_{xy}(z,0) + \int \left[ \sum_i h_i(M_{xy}(z,t)) \right] \partial t$. The two modified-Bloch equations, equations 8 and 16 are built either with the same $\frac{\partial}{\partial t}$ or the same $\partial t$, namely, the modified-Bloch equation should be built either with the same time derivative in a differential equation or the same time increment in an integral equation. When $\alpha = 1, \beta = 2$, eq 16 reduces to

$$M_{xy}(z,t) = M_{xy}(z,0) + \int_0^t \left( D \frac{\partial^2}{\partial z^2} M_{xy}(z,t) - i\gamma g(t) z M_{xy}(z,t) - \frac{M_{xy}(z,t)}{T_2} \right) d\tau, \qquad (17)$$

which is the integral type of modified-Bloch equation for normal diffusion. eq 17 is equivalent to eq 1, the regularly used differential type modified-Bloch equation for normal diffusion.[17,18,35]

PFG anomalous diffusion behaviors are significantly different from PFG normal diffusion. Figure 4 shows the comparison between free anomalous diffusion and normal diffusion. In Figure 4, the diffusion constant used for normal diffusion is $1.8 \times 10^{-11}$ m$^2$/s, while anomalous diffusion has $\alpha = 0.5, \beta = 2$ and $D_f = 9.0 \times 10^{-12}$ m$^\beta$/s$^\alpha$, which can be seen as a curvilinear diffusion; the SGP approximation is used with wavenumber equaling $\gamma g \delta = 2.6751 \times 10^5$ rad/m. Figure 4 (a) shows that the mean square displacement is a bending curve for free anomalous diffusion, while it is a straight line for normal diffusion. A similar difference can be seen in the signal attenuation curve in Figure 4 (c). These free anomalous diffusion bending curves in Figures 4 (a) and (c) could be misinterpreted as a restricted normal diffusion curve with a certain domain size, which can lead to wrong morphology structure in material studies. Figures 4 (b) and (d) plotted $\log(<z^2>)$ versus $\log(\Delta)$ and $\ln(-\ln(-S(t)))$ versus $\ln(\Delta)$ respectively, and all the curves are straight lines and the curve slopes correspond to the time derivative order parameters. No domain size is indicated in Figures 4 (b) and (d). Thus, it is important to distinguish normal diffusion and anomalous diffusion in PFG experiments. The modified-Bloch equation method is a fundamental and versatile theoretical method. It incorporates the simultaneous contributions from different processes such as diffusion, precession, and relaxation upon spin magnetization evolution. The current results in this paper only consider two important effects in PFG diffusion, the FGPW effect and spin-spin relaxation effect. Other complicated situations such as drift diffusion, boundary or initial conditions, spin-lattice relaxation, and exchange could also be considered based on the modified-Bloch equation, which requires more future research efforts.

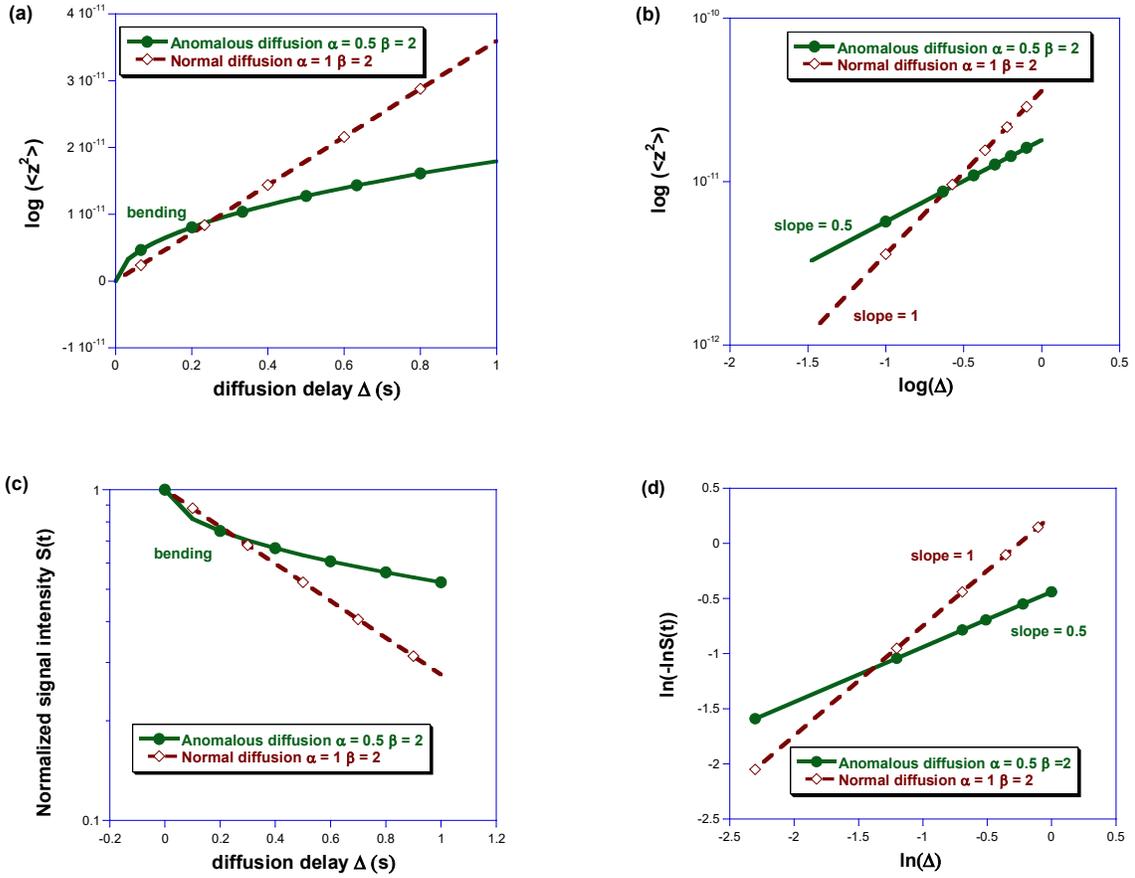

**Figure 4.** Comparison between free anomalous diffusion and normal diffusion: (a) $\log(<z^2>)$ versus $\Delta$, (b) $\log(<z^2>)$ versus $\log(\Delta)$, (c) $S(t)$ in logarithmic scale versus $\Delta$, (d) $\ln(-\ln(S(t)))$ versus $\ln(\Delta)$. The diffusion constant used for normal diffusion is $D = 1.8 \times 10^{-11}$ m$^2$/s , while anomalous diffusion has $\alpha = 0.5, \beta = 2$ and $D_f = 9.0 \times 10^{-12}$ m$^\beta$/s$^\alpha$. For the simplicity, the SGP approximation is used here with $\gamma g \delta = 2.6751 \times 10^5$ rad/m.

### References and notes

(1) Metzler, R.; Klafter, J. The random walk's guide to anomalous diffusion: a fractional dynamics approach. *Phys. Rep.* **2000**, *339*, 1−77.

(2) Klafter, J.; Sokolov, I.M. *First step in random walk. From Tools to Applications*; Oxford University Press, New York, 2011.

(3) Wyss, W. *J. Math. Phys.* **1986**, *27*, 2782−2785.

(4) Sokolov, I.M. Models of anomalous diffusion in crowded environments. *Soft Matter* **2012**, *8*, 9043−9052.

(5) Saichev, A. I.; Zaslavsky, G.M. Fractional kinetic equations: solutions and applications. *Chaos* **1997**, *7* (4), 753−764.

(6) Polanowski, P.; Sikorski, A. Simulation of molecular transport in systems containing mobile mbstacles. *J.*

*Phys.Chem. B* **2016**, *120*, 7529−7537.

(7) Kärger, J.; Pfeifer, H.; Vojta, G. Time correlation during anomalous diffusion in fractal systems and signal attenuation in NMR field-gradient spectroscopy. *Phys. Rev. A* **1988**, *37* (11), 4514–4517.

(8) Kimmich, R. *NMR: Tomography, Diffusometry, Relaxometry*; Springer-Verlag, Heidelberg, 1997.

(9) Fatkullin, N.; Kimmich, R. Theory of field-gradient NMR diffusometry of polymer segment displacements in the tube-reptation model. *Phys. Rev. E* **1995**, *52*, 3273–3276.

(10) Bennett, K.M.; Schmainda, K.M.; Bennett, R.T.; Rowe, D.B.; Lu, H.; Hyde, J.S. Characterization of continuously distributed cortical water diffusion rates with a stretched-exponential model. *Magn. Reson. Med.* **2003**, *50*, 727−734.

(11) Bennett, K.M.; Hyde, J.S.; Schmainda, K.M. Water diffusion heterogeneity index in the human brain is insensitive to the orientation of applied magnetic field gradients. *Magn. Reson. Med.* **2006**, *56*, 235−239.

(12) Magin, R.L.; Abdullah, O.; Baleanu, D.; Zhou X.J. Anomalous diffusion expressed through fractional order differential operators in the Bloch Torrey equation. *J. Magn. Reson.* **2008**, *190*, 255−270.

(13) Lin, G. An effective phase shift diffusion equation method for analysis of PFG normal and fractional diffusions. *J. Magn. Reson.* **2015**, *259*, 232−240.

(14) Hahn, E.L. Spin echoes. *Phys. Rev.* **1950**, *80*, 580−594.

(15) McCall, D. W.; Douglass, D. C.; Anderson, E. W.; Bunsenges, B. *Physik. Chem.* **1963**, *67*, 336−340.

(16) Stejskal E. O.; Tanner, J. E. *J. Chem. Phys.* **1965**, *42*, 288-292.

*(17)* Callaghan, P. *Translational Dynamics and Magnetic Resonance: Principles of Pulsed Gradient Spin Echo NMR*; Oxford University Press, 2011.

(18) Price, W.S. Pulsed-field gradient nuclear magnetic resonance as a tool for studying translational diffusion: Part 1. Basic theory. *Concepts Magn. Reson.* **1997**, *9*, 299.

(19) Price, W.S. *NMR Studies of Translational Motion: Principles and Applications*; Cambridge University Press, 2009.

(20) Grinberg, F.; Farrher, E.; Ciobanu, L.; Geffroy, F.; Bihan, D. Le; Shah, N.J. Non-gaussian diffusion imaging for enhanced contrast of brain tissue affected by ischemic stroke. *PLoS ONE* **2014**, *9* (2), e89225.

(21) Doi, M.; Edwards, S.F. *The Theory of Polymer Dynamics*; Clarendon Press, Oxford, 1986.

(22) Angelico, R.; Olsson, U.; Palazzo, G.; Ceglie, A. Surfactant curvilinear diffusion in giant wormlike micelles. *Phys. Rev. Lett.* **1998**, *81*, 2823–2826.

(23) Lin, G.; Zheng S.; Liao, X. Signal attenuation of PFG restricted anomalous diffusions in plate, sphere, and Cylinder. *J. Magn. Reson.* **2016**, *272*, 25−36.

(24) Cao, H.; Lin, G.; Jones, A. Anomalous penetrant diffusion as a probe of the local structure in a blend of poly(ethylene oxide) and poly(methyl methacrylate). *J. Polym. Sci. Part B: Polym. Phys.* **2004**, *42*, 1053–1067.

(25) Ingo, C.; Magin, R.L.; Colon-Perez, L.; Triplett, W.; Mareci, T.H. On random walks and entropy in diffusion-weighted magnetic resonance imaging studies of neural tissue. *Magn. Reson. Med.* **2014**, *71*, 617.

(26) Lin, G. Instantaneous signal attenuation method for analysis of PFG fractional Diffusions. *J. Magn. Reson.* **2016**, *269*, 36−49.

(27) Lin, G. Analyzing signal attenuation in PFG anomalous diffusion via a modified gaussian approximation based

on fractal derivative. *Physica. A* **2017**, *467*, 277−288.

(28) Lin, G. Analyzing signal attenuation in PFG anomalous diffusion via a non-gaussian phase distribution approximation approach by fractional derivatives. *J. Chem. Phys.* **2016**, *145*, No. 194202.

(29) Liang, Y.; Ye, A. Q.; Chen, W.; Gatto, R. G.; Colon-Perez, L.; Mareci, T. H.; Magin, R. L. A fractal derivative model for the haracterization of anomalous diffusion in magnetic resonance imaging. *Commun. Nonlinear Sci. Numer. Simul.* **2016**, *39*, 529−537.

(30) Liang, Y.; Chen, W.; Akpa, B. S.; Neuberger, T.; Webb, A. G.; Magin, R. L. Using spectral and cumulative spectral entropy to classify anomalous diffusion in Sephadex™ gels. *Comput. Math. Appl.* **2017**, *73* (5),765–774.

(31) Chen, W. Time space fabric underlying anomalous diffusion. *Chaos, Solitons Fract.* **2006**, *28*, 923.

(32) Chen, W.; Sun, H.; Zhang, X.; Korošak, D. Anomalous diffusion modeling by fractal and fractional derivatives. *Comput. Math. Appl.* **2010**, *5* (5), 1754–1758.

(33) Palombo, M.; Gabrielli, A.; Santis, S.D.; Cametti, C.; Ruocco, G.; Capuani, S. Spatio-temporal anomalous diffusion in heterogeneous media by nuclear magnetic resonance. *J. Chem. Phys.* **2011**, *135*, No. 034504.

(34) Bloch, F. Nuclear Induction. *Phys. Rev.* **1946**, *70*, 460−474.

(35) Torrey, H. C. Bloch Equations with Diffusion Terms. *Phys. Rev.* **1956**, *104* (3), 563－565.

(36) Hanyga, A.; Seredyńska, M. Anisotropy in high-resolution diffusion-weighted MRI and anomalous diffusion. *J. Magn. Reson.* **2012**, *220*, 85−93.

(37) Lin, G. The exact PFG signal attenuation expression based on a fractional integral modified-Bloch equation. arXiv:1706.02026; Lin, G. Fractional differential and fractional integral modified-Bloch equations for PFG anomalous diffusion and their general solutions. arXiv:1702.07116. **Author's note:** The current article draws heavily from arXiv:1702.07116. The article arXiv:1702.07116 includes three different PFG methods and two different derivative models. These three methods are the modified-Bloch equation method, the effective phase shift diffusion equation method, and the observing the signal intensity at the origin method. The two models are fractal derivative and fractional derivative model. Because the article arXiv:1702.07116 is too lengthy and there is an overabundance of new ideas, it may be difficult for many researchers to read and digest. To help the readers who is interested in the PFG anomalous diffusion, the article arXiv:1702.07116 is divided into shorter articles, each with its own focus. The current article focuses on the modified-Bloch equation based on the fractal model, while the other article, arXiv:1706.02026 focuses on the modified-Bloch equation based on the fractional derivative. Although both the two articles deal with the PFG anomalous diffusion based on modified-Bloch equation, they are significantly different. First, they use different derivative models; the fractal derivative in this article is a local operator, while the fractional derivative used in arXiv:1706.02026 is a global operator; currently, both derivative models have been extensively investigated in literature. Second, the modified-Bloch equations and their solution methods are obviously different in these two articles. Third, the obtained PFG signal attenuation in this article is a stretched exponential attenuation, while it is a Mittag-Leffler function in the arXiv:1706.02026. The current article will assist researchers to further understand the modified-Bloch equation based on the fractal derivative.

(38) Bloembergen, N.; Purcell, E. M.; Pound, R. V. Relaxation Effects in Nuclear Magnetic Resonance Absorption. *Phys. Rev.* **1948**, *73* (7), 679−746.

(39) Crank, J. *The mathematics of diffusion*; Oxford University Press; London 3$^{rd}$, 1964; p 4.

(40) Widom, A.; Chen, H.J. Fractal Brownian motion and nuclear spin echoes. *J. Phys. A: Math. Gen.* **1995**, *28*, 1243−1247.